\documentstyle[aps,prl,preprint,epsfig,floats]{revtex}

\newcommand{\expect}[1]{ {\left\langle #1 \right\rangle} }

\newcommand{\ket}[1]{ {\left| #1 \right\rangle} }

\newcommand{\fig}[1]{Figure~\ref{#1}}

\begin{document}

\draft
\preprint{}
\title{Quantum portfolios}
\author{S. M. Maurer\footnote{SMM thanks the Fannie and John Hertz Foundation for financial support.}}
\address{
Physics Department,
Stanford University,
Stanford, CA 94043
}
\author{T. Hogg and B. A. Huberman}
\address{
H.P. Sand Hill Labs,
Palo Alto, CA 94304
}
\date{\today}
\maketitle

\begin{abstract}
Quantum computation holds promise for the solution of many
intractable problems. However, since many quantum algorithms are
stochastic in nature they can only find the solution of hard
problems probabilistically. Thus the efficiency of the algorithms
has to be characterized both by the expected time to completion
{\it and} the associated variance. In order to minimize both the
running time and its uncertainty, we show that portfolios of
quantum algorithms analogous to those of finance can outperform
single algorithms when applied to the NP-complete problems such as
3-SAT.
\end{abstract}
\pacs{02.70.Lq, 03.67.Lx}

Quantum computers can, if built, efficiently solve certain
problems requiring exponential-time on a ``classical'' computer.
Quantum algorithms currently come in two main varieties: the ones
that rely on a Fourier transform~\cite{Shor1994}, and the ones
that rely on amplitude amplification~\cite{Grover1996}. Typically
the algorithms consist of a sequence of trials. After each trial a
measurement of the system produces a desired state with some
probability determined by the amplitudes of the superposition
created by the trial. Trials continue until the measurement gives
a solution, so that the number of trials and hence the running
time are random variables.

Strategies designed to optimize the performance algorithms with
variable running time (called ``Las Vegas''
algorithms~\cite{Motwani1995}), have recently gathered a lot of
interest. First, restart strategies can improve the average
performance of classical
algorithms~\cite{Alt1996,Luby1993,Gomes1997b}. Second, portfolios
analogous to those of finance can solve hard problems more
effectively than any single technique~\cite{Huberman1997a}.
Therefore an interesting question is how well these techniques
developed for classical Las Vegas algorithms can apply to quantum
computing.

Restart strategies have been examined in the context of Grover's
quantum search algorithm~\cite{Grover1996}. This algorithm
 consists of a number of trials repeated until a solution is found. Each trial has a predetermined number of
 iterations, which
  determines the probability of finding a solution. It is therefore necessary to carefully choose their
   number to optimize the running time.

Brassard et al.~\cite{Brassard1998} showed how to optimize the
expected running time of this algorithm by trading off a lower
success probability after each measurement against fewer
iterations in each trial. This amounts to a restart strategy.
Unexpectedly, the result is an algorithm that is faster on
average. Based on this result,~\cite{Huberman1997a} suggested that
there may be a more desirable trade off between the expected
search time and its variance, as in the return vs.~risk trade off
in finance. Here we perform this analysis in detail.

While the quantum restarts involved in this method are similar to
restart strategies for classical Las Vegas algorithms, there are
two distinctions. First, the number of iterations $t$ one performs
before having a chance of finding a solution is chosen ahead of
time. This isn't so in the classical case, where the algorithm
terminates as soon as a solution is found. Second, while the
probability of finding a solution increases with time in the
classical case, due to a monotonic cumulative probability
distribution $F(t)$ of finding a solution after $t$ iterations,
this isn't so in the quantum case.
 For a quantum algorithm, the amplitudes after a trial with $t$ iterations determine the probability of finding
  a solution upon a measurement. This probability is non-monotonic and periodic~\cite{Grover1996}.

Suppose there are $N$ possible states overall, and $S$ solutions
to the search problem. After $t$ iterations of amplitude
amplification the probability of finding a solution when measuring
the state is

\begin{eqnarray}
p_t & = & \sin^2((2 t + 1)\theta)
\end{eqnarray}

\noindent where $\theta$ satisfies $\sin^2 \theta=S/N$. As pointed
out in~\cite{Boyer1998}, if the number of solutions and hence
$\theta$ is known, it is possible to determine the number of
iterations $t^*$ necessary before a measurement is done such that
the result is a solution state with near absolute certainty.
Simply put, we need $(2 t^* + 1)\theta = \frac{\pi}{2}$ so that

\begin{eqnarray}
& t^* \approx \frac{\pi}{4} \sqrt{\frac{N}{S}} \approx 0.785
\sqrt{\frac{N}{S}} &
\end{eqnarray}

\noindent iterations are required in the limit of a small number of
solutions.

Since the probability of finding a solution is already high for
somewhat lower values of $t$, a lower expected running time can be
obtained by a restart strategy, i.e.~repeatedly running the
quantum algorithm with a smaller number of
iterations~\cite{Boyer1998}. In particular, with $t$ iterations
between every measurement, let $\eta_t$ be the number of
iterations required to find a solution, which in itself is a
random variable due to the variability in the number of trials.
The expected number of iterations to find the solution is

\begin{eqnarray}
\expect{ \eta_t } = \frac{t}{p_t}=\frac{t}{\sin^2((2 t + 1)\theta)}
\end{eqnarray}

\noindent In the limit of small $\theta \approx \sqrt{S/N}$, the
optimal expected waiting time is~\cite{Boyer1998}

\begin{eqnarray}
\expect{ \eta^* } & = & \frac{z}{4 \theta \sin^2(z/2)} \approx
0.690 \sqrt{N/S}.
\end{eqnarray}

Boyer et al.~\cite{Boyer1998} conclude that the restart strategy
is indeed better, since the expected number of iterations $\eta^*$
is about $12\%$ smaller. However, this result is misleading if
variance is considered as well. Specifically, a smaller number of
Grover iterations between measurements no longer seems as
attractive, since an improvement in the expected performance is
traded off against an increase in the variance in the running
time.

Put another way, one obtains a small gain in the expected waiting
time by increasing the probability that we will need at least twice
as many iterations. Quantitatively,

\begin{eqnarray}
  \expect{ \eta_t^2 } & = & \sum_{k = 1}^\infty (k t)^2 p_t (1 -
p_t)^{k - 1} \nonumber \\
& = & \frac{2 - 3 p_t + p_t^2}{(1 - p_t) p_t^2} t^2
\end{eqnarray}

\begin{figure}[tbh]\epsfxsize=\columnwidth \epsfbox{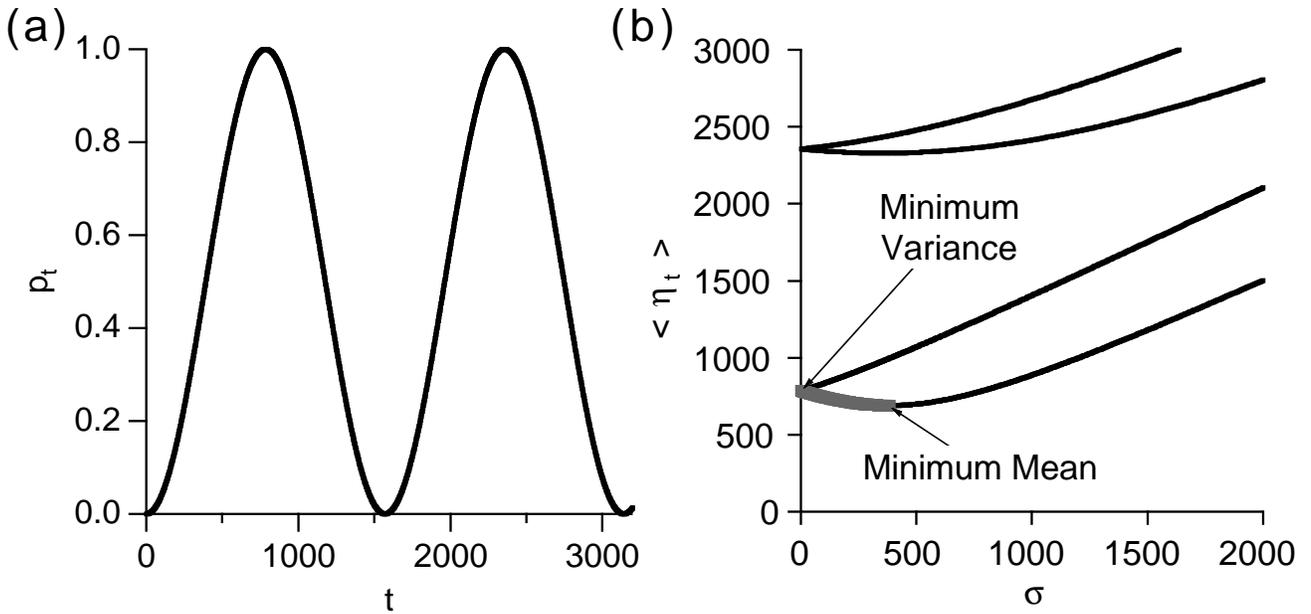}
\caption[Restart strategy in a quantum search algorithm]{(a)
Probability of measuring a solution state as a function of the
number of Grover iterations $t$. (b) Expected running time
vs.~standard deviation for a quantum search algorithm,
parameterized by the number of iterations between measurements.
The efficient frontier is emphasized. In this figure $S/N =
10^{-6}$.} \label{QuantumPortfolio}
\end{figure}

Since $\expect{ \eta_t } = t / p_t$, the standard deviation in the
number of iterations, $\sigma$, is given by $\frac{t}{p_t} \sqrt{1
- p_t}$. \fig{QuantumPortfolio} shows the ``return vs.~risk''
curve in this case. The efficient frontier contains all the points
for which the mean cannot be reduced without increasing the
variance, and the variance cannot be reduced without increasing
the mean. Any point on the efficient frontier is an acceptable
strategy, and the choice of a strategy depends on a person's
preferences. The optimum choice will depend on the application.
For example, in real-time applications, reducing variance will be
more important than reducing the expected waiting time. On the
other hand, in other domains, we may only care about the mean
performance because we average over a large number of tries.
Another possibility, common in finance, is to maximize the ratio
of return to risk, $\expect{\eta_t}/\sigma$, called the Sharpe
Ratio~\cite{Sharpe1999}. For the quantum search algorithm
presented here, the restart strategy has a Sharpe Ratio near
unity, while the strategy without restarts has a Sharpe ratio near
infinity, because $p_t$ is close to one.

This discussion also applies to any quantum algorithm whose
performance varies with the run time. This is the case of the
Grover algorithm when the number of solutions is not known a
priori~\cite{Boyer1998} and quantum adiabatic algorithms of the
type recently introduced by Farhi et al.~\cite {Farhi2001}.

Until now we considered variation in the number of iterations in
each trial. Another approach applies to quantum algorithms with
fixed trial lengths but with a variety of parameters that must be
set prior to starting each trial. For a given problem instance,
the choice of these parameters determines the probability for
success of each trial. For example, instead of just adjusting
phases based on whether states are solutions, such algorithms can
use other efficiently-computable properties, as often used in
conventional heuristic methods. In general, the best parameter
choices for a given instance are not known a priori. Thus an
important issue for such ``quantum heuristics'' is how to find
choices that work well for typical problem instances encountered
in practice.

As a specific example, we consider the $k$-satisfiability
($k$-SAT) problem.  It consists of $n$ Boolean variables and $m$
clauses. A clause is a logical OR of $k$ variables, each of which
may be negated. A solution is an assignment, i.e., a value, true
or false, for each variable, satisfying all the clauses. An
assignment is said to conflict with any clause it doesn't satisfy.
An example 2-SAT problem instance with 3 variables and 2 clauses
is $v_1$ OR (NOT $v_2$) and $v_2$ OR $v_3$, which has 4 solutions,
e.g., $v_1={\rm false}$, $v_2={\rm false}$ and $v_3={\rm true}$.
For $k\ge 3$, $k$-SAT is NP-complete~\cite{garey79}. To consider
typical rather than worst-case behavior, we focus on random 3-SAT,
in which each clause is selected randomly, and take $m=4.25 n$,
which gives a high concentration of hard instances.

For a quantum algorithm~\cite{Hogg2000}, we use a superposition of
all $N = 2^n$ assignments, adjust phases based on the number of
conflicts in each state and mix amplitudes among states based on
their Hamming distances. The algorithm's performance depends on
how well the phase adjustments match the structure of the
particular instance, which is not known a priori. One approach
finds phase choices that work well on average for random 3-SAT
problems~\cite{Hogg2000}, and gives better performance than
amplitude amplification, which ignores problem structure.

However, phase adjustments that work well on average do not work
well for all instances. Hence we can improve the performance by
using a variety of choices, thereby reducing the chance of encountering instances that perform particularly poorly for any single choice of algorithm parameters. Such a ``portfolio'' of choices can reduce both the mean and variance of the time to find a solution.

In the simplest portfolio approach, we
use different phase adjustments in each trial, instead of using the same ones
every time. As we will show, this portfolio method is
guaranteed to perform better than a fixed strategy, provided we
have no other information concerning the performance of different
choices.

To quantify this improvement, consider a variety of phase choices for a given instance. Each choice gives a particular success probability $p$ for a single trial. Let $f(p)$ be the distribution of success probabilities when selecting phase choices from among a prespecified set of possibilities.

The default strategy simply picks a single choice for the phase
adjustments to use for every trial. In this case, the expected
running time is $\expect{\frac{1}{p}}$, while the variance is
$\expect{\frac{1-p}{p^2}} + \left( \expect{\frac{1}{p^2}} -
\expect{\frac{1}{p}}^2 \right)$

Now assume that we randomly chose a different set of phases every
time. Then the probability of success for a trial is simply
$\expect{p}$. It follows that the mean and standard deviation are given by
$\frac{1}{\expect{p}}$ and $\frac{1}{\expect{p}} \sqrt{1 - \expect{p}}$, respectively.

Instead of using different choices for each trial, we can also use a ``quantum portfolio'' of our quantum
algorithms. That is, all choices are evaluated simultaneously in superposition by using additional qubits to specify the particular phase choice.

For simplicity, consider such a portfolio of only two
algorithms, i.e., choices for phases. A single qubit $\ket{a} =
\alpha \ket{+} + \beta \ket{-}$ will determine which of the two
algorithms is to be used.


Suppose the first choice gives amplitudes $c_i$ for state $i$ and
the second gives $d_i$, when run individually.  Then the result
for the quantum portfolio is
\begin{equation}
\alpha \sum_i c_i \ket{i,+} + \beta \sum_i d_i \ket{i,-}
\label{portfoliostate}
\end{equation}

\noindent The probability of measuring a solution state is

\begin{eqnarray}
p & = & |\alpha|^2 \sum_{i \in A} |c_i|^2 + |\beta|^2 \sum_{i \in A} |d_i|^2
\end{eqnarray}

\noindent This probability is just a weighted sum of the
individual success probabilities of each algorithm on its own.
This discussion generalizes to combinations of additional phase
choices. As a result, quantum portfolios of quantum algorithms,
and mixed quantum algorithms, in which we randomly use a different
quantum algorithm after each iteration, are equivalent, since they
both result in the same probability of success at every
measurement.

The surprising result is that such portfolios will complete faster on
average that a computation with a fixed set of parameters. This is
because

\begin{eqnarray}
\expect{\frac{1}{p}} \geq \frac{1}{\expect{p}}
\end{eqnarray}

\noindent always holds, since $p$ is strictly positive. Equality only
occurs for a trivial probability distribution with only one outcome.
This can be easily shown using the Schwarz inequality
\cite{Feller1971}:

\begin{eqnarray}
\expect{A B}^2 \leq \expect{A^2} \expect{B^2},
\end{eqnarray}

\noindent for which equality holds only if there is a linear
combination $a A + b B$ that is equal to zero with unit probability.

\begin{figure}[tbh]\epsfxsize=\columnwidth \epsfbox{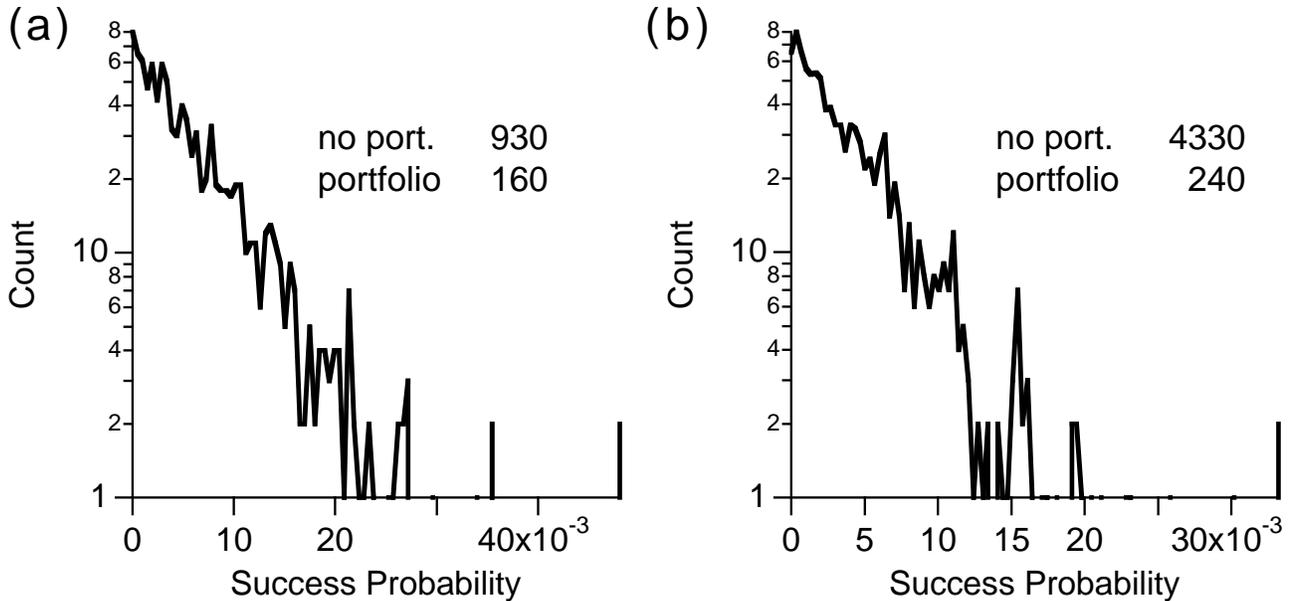}
\caption[Portfolio of unoptimized phases]{Distribution of success
probabilities upon measurement of the quantum state for randomly
chosen phases, for two different SAT instances with $n = 8$. The
numbers in the panel show the expected number of iterations with
and without restarts.} \label{Unoptimized}
\end{figure}

The performance improvement obtained with such a strategy is shown
in \fig{Unoptimized}, in which we randomly chose a different set
of phases to use for each quantum computation between
measurements. The histogram shows a substantial probability of
choosing sets of phases with near zero probability of finding a
solution after each measurement. This leads to a large expected
waiting time if a portfolio isn't used.

\begin{figure}[tbh]\epsfxsize=\columnwidth \epsfbox{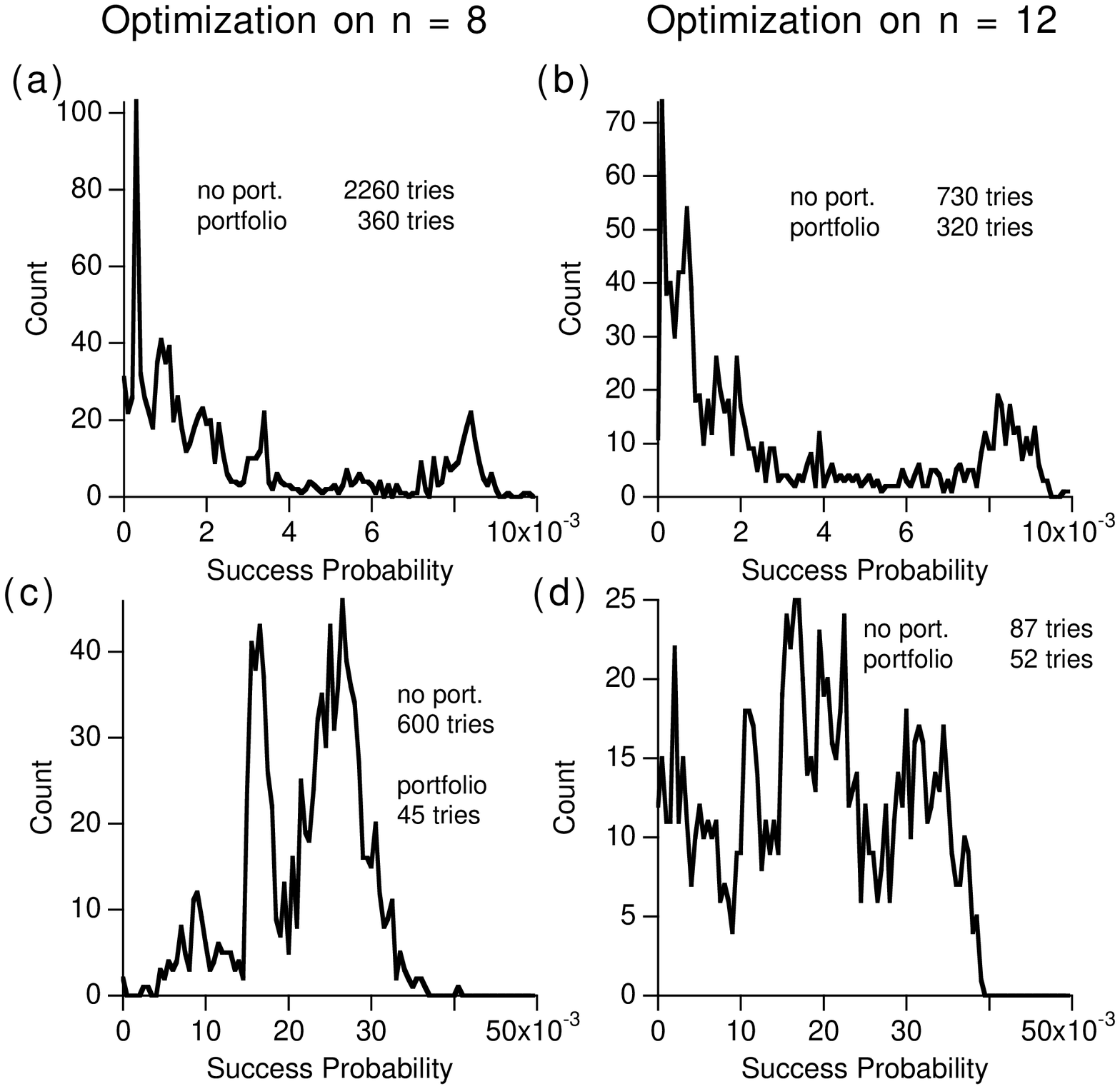}
\caption[Portfolio of optimized phases]{Distribution of success
probabilities upon measurement of the quantum state for two 3-SAT
problems with 20 variables. (a),(b) and (c),(d) correspond to two different problem instances. (a),(c) Random choice of phases optimized on instances with 8 variables. (b),(d) Random choice of phases optimized on instances with 12 variables. The expected number of measurements until a solution is found (based on the histograms) is indicated both for a single choice of parameters and a ``portfolio'' of parameters.}
\label{Optimized}
\end{figure}

In practice, one would not use entirely random phases since their
typical performance is rather low. Rather, one would use phases
that are already known to work well for similar problems, e.g.,
obtained by optimizing the choices for a sample of similar problem
instances. Optimizing choices for these sample instances multiple
times, starting from a variety of initial choices, gives a set of
phase choices that can be expected to perform much better than
random choices on new instances drawn from the same problem
ensemble as the training sample.

Since such optimization can be computationally demanding, a more
feasible scenario performs this optimization on a large sample of
small problems and then applies the resulting choices to new,
larger instances. Thus algorithms are optimized on easily solvable
instances, and applied to more difficult problems that have not
been solved before.

For random $k$-SAT, the ratio of clauses to variables, $m/n$,
characterizes the concentration of difficult instances so a
reasonable scaling approach optimizes phases for small problems
and then applies those to larger problems with the same ratio.
\fig{Optimized} shows the performance improvement of such a
strategy. Here the set of phase choices available to select was
created by optimizing for instances of 3-SAT with 8 and 12
variables and clause to variable ratio, $m/n$, of 4.25. These
optimized phases were used to solve larger instances, with 20
variables. In this case, the advantages of a portfolio strategy
are less dramatic than in \fig{Unoptimized}, but still quite
impressive.

As a further observation, phases optimized for $n = 12$ perform
better on the $n = 20$ instances than those optimized for $n = 8$.
Thus, the benefit of portfolios increases with the difference in
size between those used to select phase choices and the instances
of interest.

In the preceding, we considered a simple form of quantum
portfolios, in which all available phase choices are evaluated
simultaneously and independently for each trial. As with other
quantum algorithms operating between measurements, such trials can
be described by a single operator on the state in
Equation~\ref{portfoliostate}. Since this operator does not itself
involve a measurement, it can be used with amplitude
amplification~\cite{Boyer1998}. Thus while classical and quantum
portfolios of quantum algorithms are equivalent when run as a
series of trials, each ending with a measurement, the quantum
portfolio allows a simple further improvement by combining it with
amplitude amplification. That is, the expected performance
improves from $1/\expect{p}$ to $O(1/\sqrt{\expect{p}})$.

Thus the advantages of classical portfolios can be improved on
even further by a truly quantum portfolio. More generally, as with
combining different conventional heuristics, we could consider
operators within each trial that mix amplitude among the different
quantum algorithms. Such operations provide a broader range of
possible techniques than available with classical portfolios,
though it remains to be seen whether such extensions give
significant additional improvements.


\end{document}